# Design of an Efficient, Ease-of-use and Affordable Artificial Intelligence based Nucleic Acid Amplification Diagnosis Technology for Tuberculosis and Multi-drug Resistant Tuberculosis

Arastu Sharma[1], Rakesh Jain[2]

1. Centre for photonics, University of Cambridge, UK
2. Department Of Engineering, Suresh Gyan Vihar University, Jaipur, India

**Abstract**

Current technologies that facilitate diagnosis for simultaneous detection of Mycobacterium tuberculosis and its resistance to first-line anti-tuberculosis drugs (Isoniazid and Rifampicim) are designed for lab-based settings and are unaffordable for large scale testing implementations. The suitability of a TB diagnosis instrument, generally required in low-resource settings, to be implementable in point-of-care last mile public health centres depends on manufacturing cost, ease-of-use, automation and portability. This paper discusses a portable, low-cost, machine learning automated Nucleic acid amplification testing (NAAT) device that employs the use of a smartphone-based fluorescence detection using novel image processing and chromaticity detection algorithms. To test the instrument, real time polymerase chain reaction (qPCR) experiment on cDNA dilution spanning over two concentrations (40 ng/uL and 200 ng/uL) was performed and sensitive detection of multiplexed positive control assay was verified.

**Introduction**

Nucleic acid amplification testing (NAAT) enables highly sensitive detection of infectious diseases and forms the primary method of amplifying and detecting RNA and DNA targets. Polymerase Chain reaction (PCR) is a key NAAT process and can be performed in several different methodologies including nested (nPCR), real time reverse transcription (RT-PCR), loop mediated isothermal amplification (LAMP), and quantitative nucleic acid sequence-based amplification (QT-NASBA).[1] NAATs technologies differ depending on the requisite equipment, temperature profiles, processing time and usability, but primarily are based on two steps - involving nucleic acid extraction and followed by amplification process with either varying temperature cycle or single temperature. In modern PCR processes, quantification of the diagnosis is done by incorporating a fluorescent label to the reaction which allows the imaging device to monitor continuous change in the intensity of the amplification resulting in real-time solution as well as on comparison with titrated controls produces a diagnosis. [2,3]

Technological advancements in PCR devices have been limited to laboratory-based instrumentations. With different applications in mind in the design process, real time PCR instruments lack the required flexibility and potential. Several research studies have pursued development of low-cost solutions [4] but have unfulfilled gaps for applications in point-of-care (POC) last mile diagnosis devices. POC testing requires ease of use functioning with low resource settings to increase accessibility to diagnosis. Previous work in similar smartphone-based fluorescence detection devices lack in the ease-of-use as it is required for the user to select the region of interest.

Conventional RT-PCR requires the need of a thermocycler to perform nucleic acid amplification tests (NAAT) that is required to perform DNA amplification tests for SARS-CoV-2 or Tuberculosis. The fact that SARS-CoV-2, common cold virus and several bacterial infections yield similar symptoms, including cough, fever and rash, complicates differential diagnosis. Typically multiplexed quantitative reverse-transcription polymerase chain reaction (qRT-PCR), generally comprise bulky and complex peripheral components including expensive and power-hungry thermal cyclers (for rapidly heating and cooling samples) with built-in fluorimeter units (for performing real-time fluorescence detection).[5] A solution towards more compact, portable, and inexpensive field deployable systems is the need of the hour. Access to low-cost instrumentation could enable faster diagnosis and help in routine monitoring of spread of COVID-19 in all zones.

The device is based on a CMOS photodiode fluorescence detector, a silicon heater-based heating element as well as an ITO based heating lid. The instrument contains 16 wells and supports sample volumes between 10 – 50 µL per reaction, with a maximum ramp rate of 6 °C/s in heating and 3°C/s in cooling.  the system's heating and cooling efficacy results in rapid cycling and leads to assay completion in as little as 20 minutes. Multiple such devices can be connected also together in low-resource settings to increase the number of samples being analysed at a time; all the data will be stored together due to the smartphone cloud connectivity.

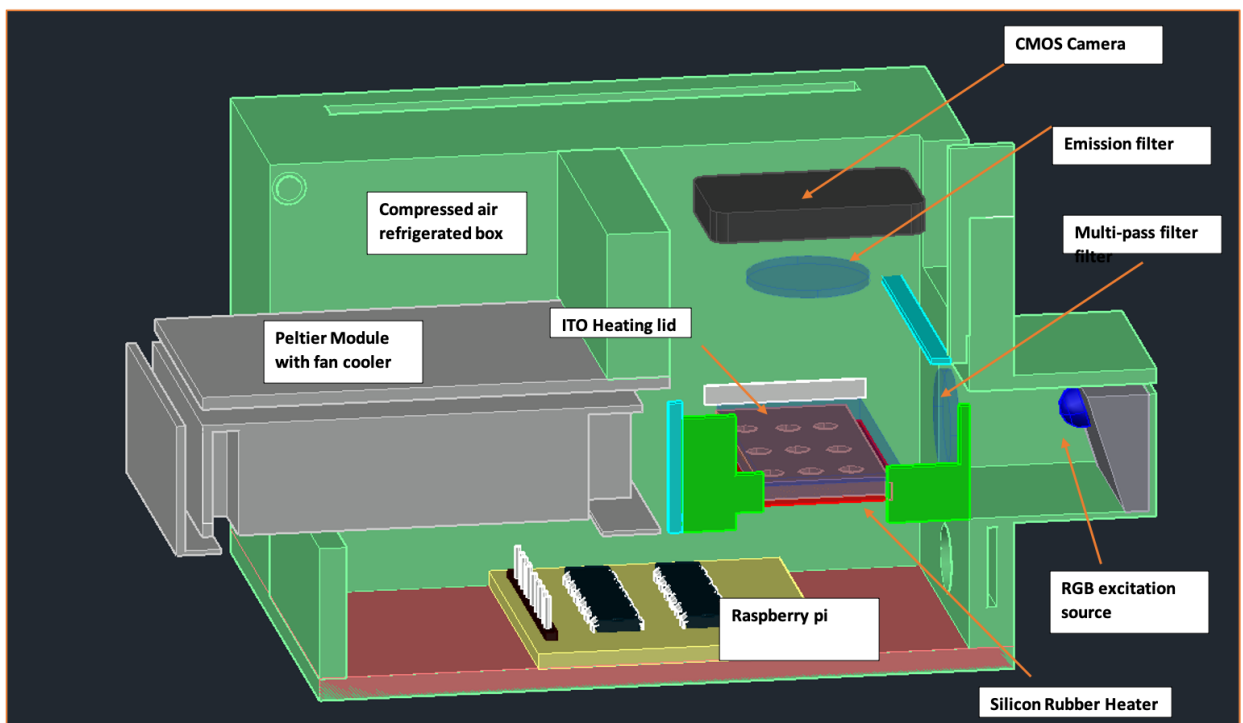

Fig 1. 3D Model render of the proposed device with cross sectional view, it covers all description of the heating and cooling elements as well as optical imaging system of the device

**Instrumentation and Methods:**

Since early demonstration of PCR instruments, very few updates have been conducted on the design process. Portable PCR instruments have been demonstrated in [5-10], but lack in sensitivity and integration processes which leaves technological gaps in realisation. The basic building blocks of the device is a i) thermocycler heating box, ii) an optical imaging CMOS camera & iii) fluorescence filters.

**Thermocycler box**

A highly flexible thermocycler box is developed with advanced heating and cooling element that speed-up the ramp cycling process. The base of the primary heating element is a silicon rubber based flexible heater that covers the copper isothermal plate from all sides. Silicon rubber provides a versatile insulation for temperatures up to 500°F (260°C). The heating device functions at the measured watt density of 20 W/in2 (3.1 W/cm2), dependent upon application temperature. The thickness of the heater is 0.018 in. (0.5 mm) and is manufactured using a foil element. The cooling element involves several elements for speeding up the reverse ramp cycle including a Peltier thermoelectric device, refrigerator gas compressor and a computer fan blower with the fan nozzle faced directly onto the heating element. The temperature was monitored in real-time with a NTC thermistor connected to an analogue-to-digital converter. The accuracy of the temperature control is measured based on the calculating the difference of set points and average measured temperature.

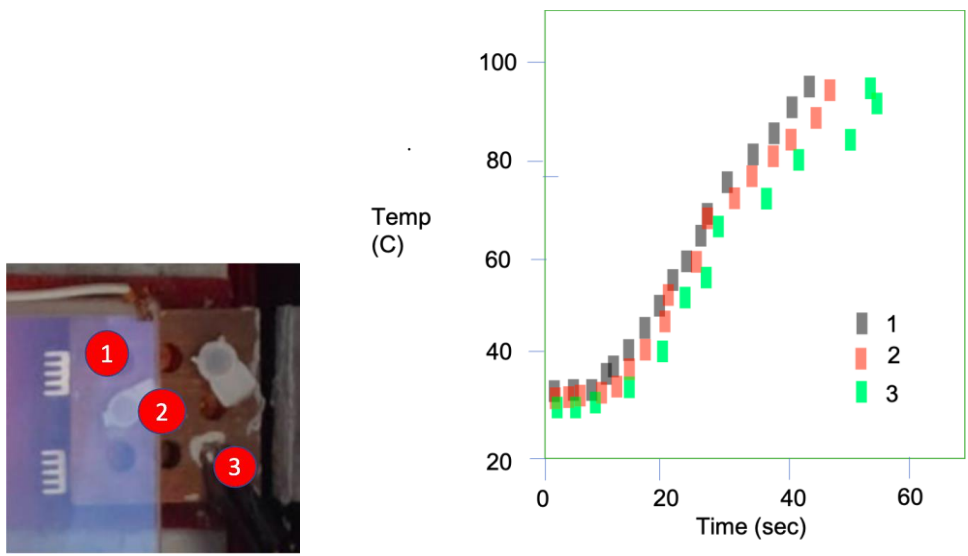

Figure 2. Temperature Characterisation of the copper base is shown to shown as the temperature measured on the dots represented in 2.a) are mapped against time in 2.b). The figure shows isothermal behaviour over the whole testing surface.

**Artificial Intelligent PID control**

The primary controlled system is based out of a Raspberry Pi 4 device. The analogue devices are controlled using ADC converters connected GPIO pins on the raspberry device. A python based artificial intelligence PID controlled device was designed. The machine learning code plots the change in ramp cycles with the power supplied to three different heating and cooling elements used in the device, including the silicon rubber heater, the refrigerator compressor, the Peltier module and the blower. The positive reinforcement is setup on the feedback loop with the aim of reducing the ramp cycle for both increasing and decreasing the temperature in the thermocycler. The proportional integration derivative (PID) control parameters were designed based on strategy which uses power delivery with pulse width modulation (PWM) to heating and cooling elements.

**Optical Setup and Analysis**

For an optical imaging setup to work perpendicular to the micro-PCR tubes, an ITO based heating lid is used, which is a transparent heat conductor. A 3W RGB tricolour LED source is placed 20° over the horizontal plane, with a triple-bandpass filter. The imaging system is based on a CMOS camera (8.08MP RGB-CMOS sensor) of an android smartphone and is controlled using an app developed in Python and android platform. The digital zoom abilities of the current CMOS technology provide a 5x magnification using the developed affine to the technology and without any pixel degradation. Images are obtained using combination of excitation and emission filters (fluorescence filter) placed with the light source and CMOS camera respectively.

The communication and control between the smartphone and Raspberry pi happens using a Bluetooth module and separate function written in python library. The app provides the ease-of-use innovation in the device with a very simplistic custom designed Graphic user interface (GUI), which comprises of the digital control screen, automated sample selected screen and Colorimetric-Luminance image analysis screen. Ramp cycles of the thermocycler and other electronic control is provided on selection of already setup cycle information of several assays of Sars-Cov-2 and Tuberculosis. With an additional possibility to add cycling settings with a simple android update, the device is adaptable to conduct all types of NAAT amplification techniques including qPCR, real time and LAMP. Users have the choice of manual and automated imaging settings. The manual settings give options to make targeted area selection, while the automated AI imaging process acquired the image based on computer vision techniques.

**Post processing after AI image acquisition**

The images were captured through a 8 mega-pixel CMOS camera and sent to the smartphone for post processing analysis. Images are taken every 15 seconds and are saved in a bit mat data file format that contains 4 bits per each pixel, containing in the set values for red, blue, green and alpha pixels, corresponding to standard values of RGB colour space. The first complication of the process is identifying the region of interest (ROI) using computer vision image extraction techniques. Through a special algorithm, we the exposed tube area is matched with a reference image model, that automatically maps the specific tube area which can show fluorescence or be empty tube structure. The image is mapped into smaller pixel boxes and then Image hashing as well as template matching algorithm [6] is performed to identify similar pattern of pixel gradient on the plane. Template matching algorithm is explained further in figure 3. Collected image pictures from generalised RTPCR are matched to the current test images to focus on the right crop.

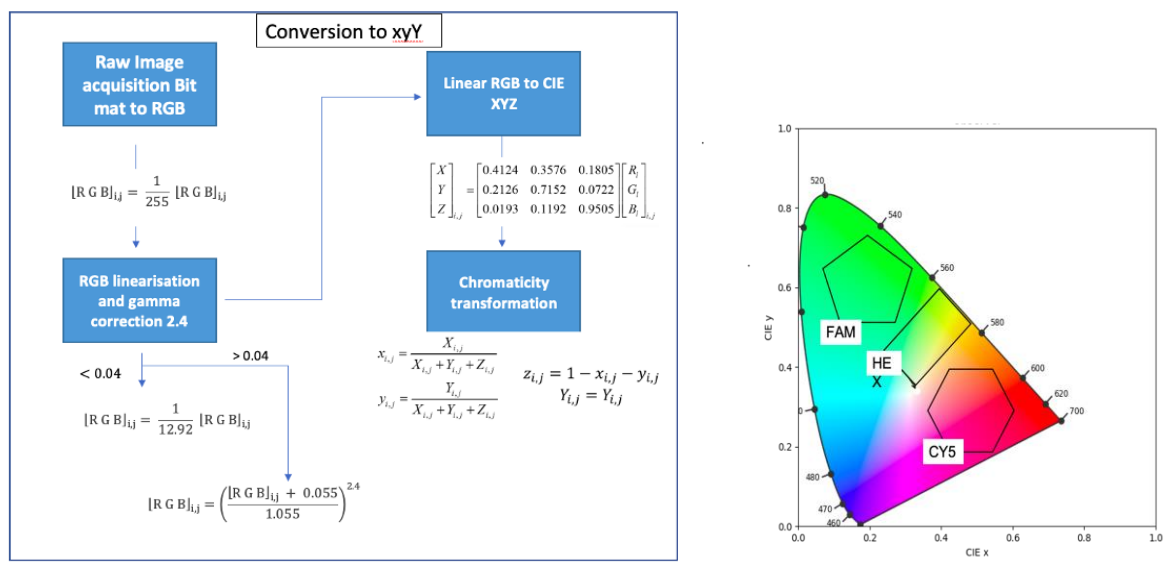

Figure 3. a) Shows the algorithm for image processing as RGB image data set is converted to be represented on a CIE chromaticity diagram [14], b) Die islands are shows as processed in the algorithm

After the image acquisition, the alpha values are ignored as its constant for all, and series are transformations are implemented on the cropped images. First a Gamma transform is implemented on with a value of 2.4 to cancel the non-linearity intensity scale of CMOS sensor [14]. A series of transformations in python imaging is processed based on the chart shown in figure 3, to convert raw RGB data to CIE xyY colour space. Python module library "colour-science" was used to plot the CIE 1951 chromaticity diagram, registered under the open-source license [8]. Visualisation of the chromaticity is done by using chromaticity values mapped on 2d chromaticity diagram on a predefined corresponding island of different types of emission spectrums produced by fluorophores, namely, FAM, HEX, ROX and CY5.[14] The island boundaries are dependent on the wavelength of emission as well as the colour gamut as shown in figure 3.

**Artificial Intelligent fluorescence detection and image automation**

The paper indicates Artificial intelligence-based detection of colour as a method of increasing sensitivity in low resource areas with lighting issues as well as increasing rate of detection by predicting quantisation. Data set was acquired from covid testing facility with both Negative Test Control (NTC) and Positive Test Control (PTC) for every 30 seconds of run. The data on chrominance of cropped images is compared to reference images of NTC and PTC, to calculate the SAD – sum of absolute difference for all the pixels. This process was used to fine tune the SAD threshold values which is fixed for the assay. In reality, SAD models don't have a general threshold and thus machine learning models are required in increasing sensitivity of detection using correct classification of colour.

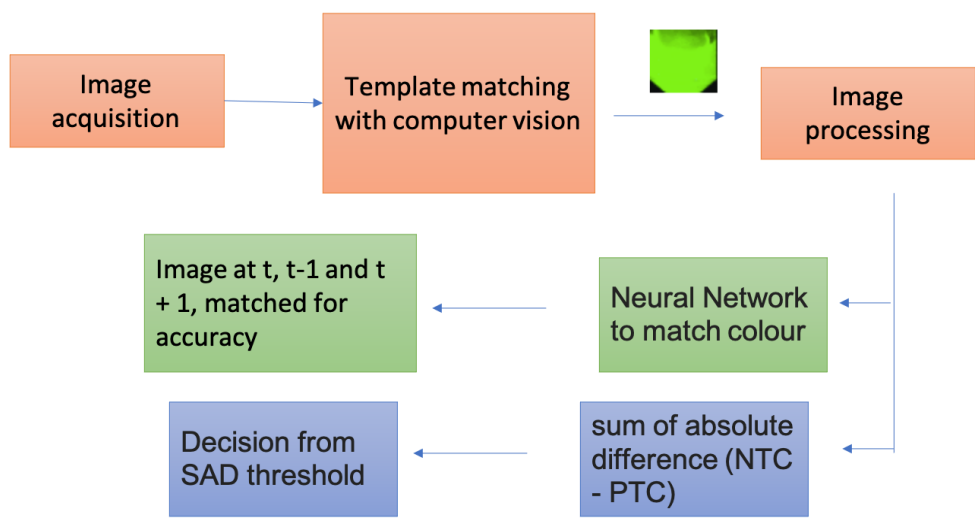

Figure 4 Shows the two methods considered for image acquisition and sensitive detection, data proves that trained neural network method is highly successful

We collected 800 cropped images distributed into groups depending on better light. Most of the data was used for training the neural network, while 50 images were exploited to see change in learning after each epoch. The machine app currently involves a method of data augmentation and in the process the sensitivity will increase with higher testing using the device using real time data sets. Data augmentation capabilities in the device make it possible to add variety to the data, such as random brightness, tilt, and shift by pixels; and therefore, make the device less susceptible to the environment. The neural network convergence was observed to happen after 11 epochs, which can be further improved with a larger data set. Neural network model is used to identify the colour with reduced discrepancies as images from t, t-1 and t+1 is matched together, and then compared to NTC and PTC.

**Experiments and Discussion:**

The instrument was designed to perform real-time PCR on samples Anyplex™II MTB/MDR assay was used to experiment detection on the device. The assay uses multiplexed die for detection of Mycobacterium tuberculosis and drug resistance to Isoniazid (related gene katG, inhA) and Rifampicin (related gene rpoB). The increase in concentration of DNA sample should increase with each cycle and should be visualised by increase in fluorescent amplitude. The optical setup of our device institutes multi band pass filters thus supports simultaneous multiplexed detection of die.

To test the efficiency of our approach, we plot the observed Fluorescence emission intensity against temperature cycle profile and cycle number. Two different values of DNA concentrations, 200 ng/uL and 40 ng/uL are used to understand the efficiency of the device. The diagram plots fluorescence emission at each time points normalised to fluorescence emission of baseline. Cycle threshold (CT) values of 17 and 24 are observed for 200 ng/uL and 40 ng/uL for DNA concentrations, it is observed that the CT decreases with increasing the concentration as well as that negative control test (no target) remains zero for all the cycles.

Duplex detection of presence of Mycobacterium tuberculosis and resistance to Isoniazid (4 mutations in katG) was performed with detection targeted in FAM and CY5 regions. Intact TB DNA samples are used targets in this experiment. Samples were illuminated with RGB LED based lights and filtered through fluorescent excitation filters to get the targeted wavelengths. The analysed images are mapped into pixelated targets and using the template matching algorithms are automatically detected. Further, the detected colours are plotted on the fluorophore emissions islands on the CIE diagrams. CIE Y represents ratio of the positive and negative for the luminance value and is plotted for FAM and CY5 detections for our experiment on the figure 6. The portable device can detect multiplexed die detection with similar sensititvity to a benchtop RT-PCR and thus can be suggested for alternatives of such devices in low-resource settings with further improvements.

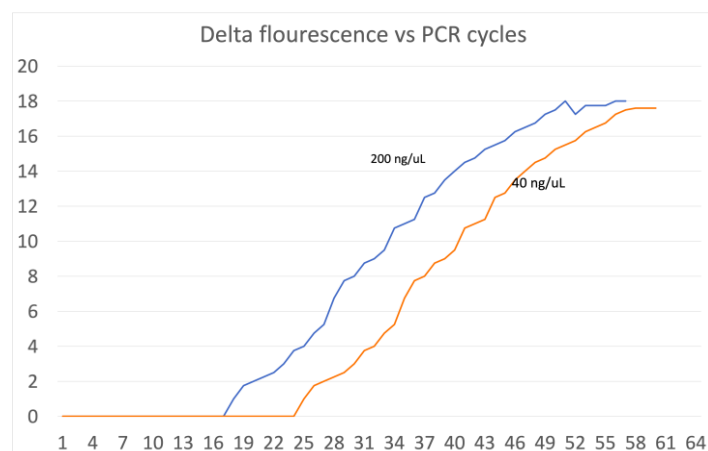

Figure 5 Real Time PCR fluorescence intensity curve against temperature cycles to prove device sensitivity

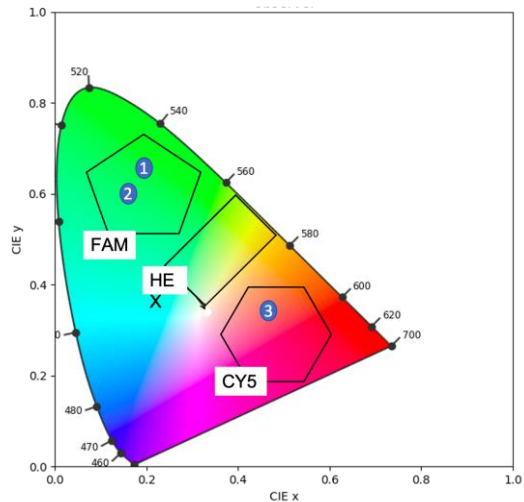

Figure 6 Multiplex detection of Anyplex™II MTB/MDR assay is plotted for Mycobacterium tuberculosis and resistance to Isoniazid (4 mutations in katG)

## Conclusions

The primary aim of this point-of-care last mile solution is to reach to low resource settings and provide a highly sensitive alternative to lab based nucleic amplification infectious disease detection. In this paper, we demonstrate a quantitative real time NAAT detection technology with multi-dye detection capability and adaptable thermocycler for both qPCR and LAMP techniques (isothermal NAA assays). The advanced AI implementation in use for accuracy of the PID system as well as the optical imaging device makes it the first NAAT device to use these technologies to the best of our knowledge. This device yet lags in performance of a commercial thermocycler, measured against CT efficiency. Efficiency can be improved with various changes in design parameters, number of reagents used in a single PCR reaction as well as developed of a targeted assay for the device, can provide a much better solution.

This would allow for widespread use of these technologies in low-resource settings and contribute to the diagnosis, treatment, and control of certain diseases. To control the thermocycler and the optical setup, we utilized a Raspberry Pi and Python, some of the most popular computer and opensource platforms in the world that could expedite the adoption of these technologies in low-resource settings. To make the instrument user-friendly, we used a tablet with a display showing real-time temperature and the change in fluorescence intensity inside the wells of the device as the PCR reaction unfolds. The instrument allowed us to perform PCR on a series of dilutions of DNA and obtain quantitative measurements of the initial amount of genetic material in the samples.

## Further work:

The team is expanding into microfluidics to explore development of Lab-on-a-Chip (LOC) based-devices to integrate multiple laboratory functions on a single chip of only a few square millimeters to a few square centimetres in size. The idea is to provide miniaturized, automated, integrated and parallelized chemical and/or biological analyses of picolitres of sample with a high performance at high speed and least cost.

## Acknowledgment

Dr. Sandeep K Shrivastava, Founder Head to CIRD and Head Dept of Allergy and Molecular Diagnostics of Dr. B. Lal Inst. of Biotechnology and his team for providing the data set for machine learning analysis as well conducting experiments in their lab.